# Substrate inhibition imposes fitness penalty at high protein stability

Bharat V. Adkar[a], Sanchari Bhattacharyya[a], Amy I. Gilson[a], Wenli Zhang[a,b], Eugene I. Shakhnovich[a,1]

[a]Department of Chemistry and Chemical Biology, Harvard University, 12 Oxford St., Cambridge, MA 02138, USA
[b]State Key Laboratory of Food Science and Technology, Jiangnan University, Wuxi 214122, China

[1]Correspondence should be addressed to E.I.S. (shakhnovich@chemistry.harvard.edu)

# Abstract

Proteins are only moderately stable. It has long been debated whether this narrow range of stabilities is solely a result of neutral drift towards lower stability or purifying selection against excess stability is also at work — for which no experimental evidence was found so far. Here we show that mutations outside the active site in the essential *E. coli* enzyme adenylate kinase result in stability-dependent increase in substrate inhibition by AMP, thereby impairing overall enzyme activity at high stability. Such inhibition caused substantial fitness defects not only in the presence of excess substrate but also under physiological conditions. In the latter case, substrate inhibition caused differential accumulation of AMP in the stationary phase for the inhibition prone mutants. Further, we show that changes in flux through Adk could accurately describe the variation in fitness effects. Taken together, these data suggest that selection against substrate inhibition and hence excess stability may have resulted in a narrow range of optimal stability observed for modern proteins.

# Introduction

Most proteins (except IDPs) must be sufficiently stable to fold to a native 3D structure, resist thermal fluctuations and proteolytic degradation in the cell and carry out function. Hence selection for protein folding stability must have been an important parameter during evolution. Naïvely this suggests that proteins would continuously evolve towards higher thermostability. In reality, however, this is not the case, and in fact most natural proteins are only moderately stable, with $\Delta G_{folding}$ in the range of −5 to −10 kcal/mol (1-3). Origins of such a narrow range of stabilities has intrigued researchers for long. Theoretical approaches which addressed this issue have employed evolutionary simulations, where studies have shown that on a protein folding driven thermodynamic fitness landscape, selection for folding stability need not result in highly stable proteins (3-5). In the regime of unstable proteins, selection for folding stability would lead to fixation of predominantly stabilizing mutations. On the other hand, in the regime of stable proteins, both stabilizing as well as destabilizing mutations have a very low selection coefficient, and hence have a low probability of fixation. However, since the supply of mutations is largely destabilizing, this results in more destabilizing mutations being fixed in the population (4-8). At some intermediate value of folding stability, mutation-selection balance happens, where stabilizing and destabilizing mutations have equal probability of getting fixed, thereby giving rise to the observation that proteins are marginally stable.

A contrary hypothesis has also been provided which states that marginal stability is the result of a fitness penalty at very high protein stability (9, 10). In other words, there is an optimal stability of proteins, beyond which on both sides fitness drops. It was hypothesized that drop in fitness at high stability is due to loss of protein flexibility that is important for its activity, resistance to proteolytic degradation, etc. At the heart of this fitness penalty, lies the concept of stability-activity tradeoff (11). Indeed, directed evolution experiments that aim to improve protein stability with no constraint on its function often lead to mutations in the active site and subsequent loss in activity (12). Moreover, specific substitutions in the active site of a protein often lead to stabilization with loss of activity (11, 13-15). This observation can be partly attributed to the fact that most substitutions in a protein are deleterious (16, 17). However, such trade-off can also be real as nature had to compromise protein stability while carving out an active site on a stable 3D scaffold, and active sites often have unfavorable conformations like buried polar amino acids or like-charges proximal to each other, etc. However, for substitutions outside the active site, such trade-off has not been demonstrated convincingly. Instead a positive correlation between stability and activity was found in one case (15). Thermophilic counterparts of mesophilic enzymes present an interesting case to explore stability-activity trade-off, however such studies have also yielded contradictory results. HD exchange experiments showed that thermophilic 3-isopropylmalate dehydrogenase was significantly more rigid at room temperature than the *E. coli* enzyme, with concomitant loss in activity (18). On the contrary, Nyugen *et al* successfully reconstructed a thermophilic ancestral adenylate kinase that was both thermostable as well as had comparable activity as the mesophilic enzyme at lower temperature (19).

To address whether high stability impairs activity with possible consequences for bacterial fitness, we used adenylate kinase (Adk) as a model protein. Adk is a reversible enzyme that inter-converts among adenylate currencies (ATP, ADP and AMP), and is essential in all forms of life. It undergoes large conformational fluctuations during catalysis (20-22), and therefore the effect of global protein stability on its activity is particularly interesting. In a previous study, we found that destabilizing mutations in Adk preferentially modulate *E. coli* lag times through changes in Adk catalytic capacity $\left(abundance \times k_{cat}/K_M\right)$ (23). In this study, we introduced stabilizing mutations in Adk outside its active site. We found a positive correlation between the conventional activity parameter $k_{cat}/K_M$ and stability, implying no trade-off, as was posited in other studies. Interestingly, we uncover a positive correlation between stability and substrate inhibition by AMP. This molecular-level effect has strong implications for physiology of *E. coli*. When placed on the genome, such stabilized variants of Adk exhibit extended lag times during growth. Furthermore, in the presence of external AMP, extension of lag times became much more pronounced along with significant drop in growth rates, which could be captured effectively using all the measurable biophysical and cellular properties of the enzyme. Our study therefore reveals a hitherto unexplored aspect of a protein's activity, substrate inhibition, that is substantially modulated by its global stability, and can potentially explain the observed moderate stability of mesophilic Adks.

## Results

**Stable mutants of Adk show increased substrate inhibition by AMP**

We engineered several stabilizing mutations at 8 different locations in *E. coli* Adk (Fig 1A, Table 1) based on two different approaches: first, using a consensus approach, where we substituted the *E. coli* amino acid with the most conserved amino acid at that location based on a multiple sequence alignment (see Methods); second, replacing by an amino acid that has been shown to stabilize Adk from *Bacillus subtilis* (24, 25) (see Methods) at equivalent position on the structure. The mutations chosen by the consensus approach were far away from the active site, *i.e.*, they were at least 8 Å away from Ap5A, an inhibitor of Adk (PDB:1ake (26)). The mutations identified using the above two approaches were single-site mutants with $\Delta T_m$ in the range of 1-6 °C. Further, we combined mutations if their $C_\alpha$-$C_\alpha$ distance on the structure were ≥10 Å (see Methods). Overall, the range of stability gains obtained for the single-site as well as multi-site mutants was 1-9 °C above WT Adk in terms of $T_m$. We measured the activity of Adk in the direction of ADP formation: $ATP + AMP \rightarrow 2ADP$. We used saturating concentration of ATP and varied concentration of AMP in reactions. The initial velocity vs AMP concentration plot for Adk deviates from the conventional Michaelis-Menten kinetics (Fig 1B). The rate of reaction decreases at high AMP concentration, exhibiting substrate inhibition. Indeed, such inhibition of Adk by AMP has been reported previously (27, 28). We used the following model of uncompetitive inhibition to quantify the effect of substrate inhibition:

$$v_0 = \frac{k_{cat}[E_0][AMP]}{K_M + [AMP]\left(1 + \frac{[AMP]}{K_I}\right)} \quad (1)$$

where $v_0$ is the initial rate of ADP formation, $[E_0]$ and $[AMP]$ are the concentrations of Adk and AMP, respectively, used in the reaction, $k_{cat}$ is the turnover rate of Adk, whereas $K_M$ and $K_I$ are the affinity- and inhibition-constants related to AMP, respectively. The $K_I$ of WT Adk at 25°C was 930 μM. Interestingly, for most of the stabilized Adk mutants, inhibition was much stronger than WT $\left(\text{lower } K_I\right)$ (representative plots for WT (blue) and a stable mutant (red) are shown in Fig 1B). To find out if substrate inhibition was related to stability of the Adk proteins, we selected a set of destabilized mutants from our previous study (23) (Table 1). Remarkably, substrate inhibition was almost completely abolished for most of the destabilized mutants (e.g. green line in Fig 1B shows the kinetic curve for destabilized mutant V106W). For a wide range of $T_m$ on either side of WT, $K_I$ values showed a strong dependence on stability $(T_m)$ of the proteins (Fig 1C, $r = -0.67$, $p = 1.1e-3$), with higher stability resulting in stronger substrate inhibition.

However, we note that mutant L82V showed strong inhibition, even though it was only marginally more stable than WT $(\Delta T_m = 0.6\ °C)$. In fact, it showed the strongest inhibition among all mutants considered in this study. This suggests that in addition to changes in stability, $K_I$ may be modulated by position specific effects, presumably through allostery. With the exclusion of L82V, the correlation between stability and substrate inhibition is much stronger ($r = -0.80$, $p = 3.3e-5$).

**Variation of enzymatic activity with stability**

Previous studies that sampled mutations outside the active-site found no trade-off between activity and stability of the protein (15, 29). However, in case of mutant Adk proteins that span ~20°C range of stability, we observed that while most destabilized proteins had $k_{cat}$ values close to WT levels, several of the stabilized mutants show a slight drop in $k_{cat}$, thereby resulting in an overall weak negative correlation with stability (Fig S1A, $r = -0.46$, $p = 1.8e-2$). We excluded all mutations that involve $16^{th}$ position from all correlations reported in this section as Q16 is very close to the active site residues ($C_\alpha$ distance to the closest active site residue is <4 Å) (Fig 1A). Contrary to $k_{cat}$, affinity towards AMP $(K_M)$ improves significantly with stability (Fig S1B, $r = -0.71$, $p = 5.2e-5$), which in turn drives the positive correlation between stability and $k_{cat}/K_M$. In other words, the enzyme becomes more efficient as it becomes more stable and this implies no trade-off between activity and stability for mutations away from active site (Fig 2A, $r = 0.68$, $p = 1.3e-4$). Similar positive correlation was reported for another enzyme, DHFR, in a previous study (15).

Interestingly, we also found that strong affinity $(K_M)$ of Adk towards AMP also results in strong inhibition by AMP $(K_I)$ (Fig 2B, $r = 0.66$, $p = 1.6e-3$). At the same time, enzyme efficiency $(k_{cat}/K_M)$ and substrate inhibition $(K_I)$ trade off (Fig 2C, $r = -0.65$, $p = 2.8e-3$).

Such trade-off is a significant effect as it implies that efficiency of stabilized forms of Adk is limited due to inhibition.

Mutant Q16F was the most stabilizing single site mutant in our study $\left(\Delta T_m = 6.5\ °C\right)$. It was selected based on a previous laboratory evolution experiment in *Bacillus subtilis* (24). However, this mutant and all combination mutants containing Q16F had very low activity as Q16 is very close to the active site (Fig 1A, red sphere and Fig 2A, red circles). These mutants therefore represent a classic case of activity-stability tradeoff at the active site. Interestingly these mutants did not show any detectable substrate inhibition in the concentration range of AMP studied (Table 1). This represents a case where extreme loss in activity leads to complete abolition of inhibition, and therefore is in line with our observation that $k_{cat}/K_M$ and $K_I$ trade-off.

**Flux through Adk explains drop in fitness**

Our biophysical studies demonstrated that stabilizing mutants exhibit strong AMP-dependent substrate inhibition. We therefore hypothesized that the inhibitory effects of AMP on the essential enzyme Adk could result in fitness defects when grown in the presence of large excess of external AMP. To that end, we engineered a selected subset of stabilized and destabilized Adk variants on the genomic copy in *E. coli* and measured fitness effects (growth rate and lag time of engineered strains) as well as intracellular abundance of the mutant Adks (Table S1) in the presence and absence of AMP. To find out the dynamic range of AMP concentrations in which the largest change in fitness effects are seen, we first measured growth curves of WT and the most inhibited mutant in this study, L82V, in a minimal media (M9) and a 0-10 mM range of external AMP concentrations. Indeed, we found that the lag time increased, for both WT and L82V, with addition of excess AMP up to ~400 μM, beyond which there was no substantial change (Fig S2). Subsequently we carried out growth experiments with all Adk variants in 0-400 μM range of AMP. Remarkably, only stabilized mutants which exhibit strong substrate inhibition (low $K_I$ values) showed an AMP-dependent drop in growth rate and increase in lag times, whereas the uninhibited mutants exhibited little-or-no effect (Fig 3A,B,C). This also shows that there is no generic toxicity due to additional AMP – the effect stems from inhibition of Adk by additional AMP in WT and some mutants. We utilized flux-dynamics theory (30) to relate the changes in fitness to changes in flux through Adk when excess AMP is present. The theory has been successfully used previously to explain fitness dependence on the activity of β-galactosidase (31), DHFR (32-34), and Adk (23). In the present case, we model the fitness dependence as follows:

$$\text{fitness}, F = a\frac{V}{b+V} \quad (2)$$

where, $V$ is the flux through Adk, $a$ is the maximum fitness when the flux is maximum, and $b$ is a constant representing background effect from all other enzymes. The flux through Adk is modeled as the rate at which AMP is converted to ADP and is related to equation (1) as follows:

$$V = Abundance \times \frac{k_{cat}[AMP]}{K_M + [AMP]\left(1 + \frac{[AMP]}{K_I}\right)} \quad (3)$$

Using measured intracellular abundances and biophysical properties of Adk variants $(k_{cat}, K_M, K_I)$, we calculated $V$ for WT and mutants at zero external AMP $(V_0)$ and under different concentrations of external AMP $(V_{AMP})$. For all $(V_0)$ calculations, we assumed intracellular AMP concentration to be 280 µM (35), while for $(V_{AMP})$, it was $280 + [AMP]$ µM. Since change in fitness is the largest for inhibition-prone mutants which in turn have low flux due to strong $K_I$ values as per Eq (3), we can assume that for such mutants $V \ll b$. In such a regime, Eq (2) simplifies to the following form:

$$\text{fitness}, F \approx a\frac{V}{b}$$

In the opposite regime of high flux through Adk when $V \geq b$ fitness becomes weakly dependent on V

$$\text{fitness}, F \approx a\left(1 - \frac{b}{V}\right)$$

which further implies that the change in fitness upon addition of AMP $\Delta F = (F_{AMP} - F_0) \propto (V_{AMP} - V_0)$ when $V \ll b$, and approximately plateaus with $V_{AMP} - V_0$ when $V \geq b$. Hence, we projected fitness components (change in growth rate and lag times) on $(V_{AMP} - V_0)$ (Fig 3D,E), and found that the change in flux upon AMP addition is well described by Eq (2) and it correlated with fitness changes very significantly with Spearman $\rho = 0.65, p = 1.5e-9$ for growth rate and $\rho = -0.78, p = 3.0e-15$ for lag time, respectively (Fig 3D,E). For mutants which show strong AMP dependent inhibition (shown in circles in Fig 3D,E), addition of AMP causes a drop in flux which is reflected in concomitant drop in fitness (decreased growth rates or increased lag times). On the other hand, mutants that lack inhibition (represented as triangles in Fig 3D,E) show an increase in flux with additional AMP. Consistent with the flux-dynamics theory Eq (2) predicts the law of diminishing returns (30, 31, 33), increase in flux beyond its native levels do not change fitness for these mutants. The mutants that do not show substrate inhibition therefore remain on the fitness plateau.

**Additional AMP leads to accumulation of adenylate metabolites in the exponential phase**

Adk is an essential enzyme that interconverts adenylate currencies in the cell. It was interesting therefore to find out what happens to levels of ATP, ADP and AMP in mutant strains and under conditions of AMP inhibition. To that end, we measured intracellular levels of relevant metabolites of a selected set of mutants in the absence and in the presence of high concentrations of external AMP during exponential phase of growth. In the absence of external AMP, levels of three adenylate metabolites in mutant strains did not differ significantly from WT (Fig 4). However, in the presence of 400 µM AMP in growth media, mutants L107I+V169E and L82V, which show strong substrate inhibition, accumulated extremely high levels of all three metabolites ATP, ADP and AMP. In contrast, destabilized mutant V106W and WT did not accumulate these metabolites even at high AMP concentration. Presumably, in the presence of high AMP concentration, the majority of the mutant Adks that have low $K_I$ remain in inhibited (bound) form, thereby not allowing the enzyme to carry out reaction in any direction and leading to accumulation of all three substrates. Previous studies have shown that accumulated AMP in yeast is often converted to IMP to prevent the slowdown of growth (36). In our study too, we observe accumulation of IMP in all four strains. The buildup is higher for L107I+V169E and L082V, as they accumulate more AMP due to inhibition.

**Physiological effect of substrate inhibition**

The data presented so far establishes conclusively that the increase in stability results in higher substrate inhibition in Adk. Such inhibition is also reflected in loss of fitness given appropriate conditions of excess substrate and therefore those 'selected' conditions can potentially limit the stability range of the protein. However, can substrate inhibition be realized under physiological conditions? This is an important evolutionary question, because if true, then there is a fitness cost due to increased stability. Brauer *et. al.* (37) showed that sudden and severe limitation of carbon source resulted in accumulation of AMP in *E. coli*. Based on this finding, we presumed that AMP might accumulate during the stationary phase too, as during this time carbon and other energy resources deplete. In such a scenario, when cells resume a new cycle of growth upon resource availability, the mutants with substrate inhibition will result in extended lag and subsequent fitness loss. To that end, we carried out metabolomics analysis of WT and L82V mutant strains during different phases of growth. Surprisingly, however, we found that all adenylate metabolites, including AMP drastically drop in the stationary phase as compared to the exponential phase (Fig 5A). The observed difference between these two experiments might arise because Brauer et.al. deprived the cells of carbon source in the exponential phase, while in our experiments all resources, including carbon, gradually decrease as a function of growth. In our study, the AMP levels in stationary phase drop to ~25% of that in exponential phase, ADP to ~15%, and ATP to almost 10% in WT. The overall pattern of drop in metabolites remains similar in L82V, the most inhibited mutant in this study. However, on a closer look, we found that relative to WT, mutant L82V contained more of all three adenylate metabolites (Fig 5B) in stationary phase, as opposed to during exponential phase where they were not substantially different (Fig 4 and Fig 5A). Specifically, AMP levels were ~1.38 times higher in L82V compared to WT. So, can higher levels of AMP in mutants relative to WT during stationary phase explain the variation in lag times at zero external AMP? To address this question, we calculated change in flux between mutant and WT Adk $(V_{mut} - V_{WT})$ using equation (3) in the following way: assuming AMP concentration during exponential phase of *E. coli* growth to be 280 μM (35), and AMP levels in stationary phase to be 25% of exponential levels (this study, Fig 5A), we consider $[AMP]_{WT}$ to be 70 μM for flux calculations. Next, we assumed all mutants to have same levels of AMP $([AMP]_{mut})$ in stationary phase and it was set to 96 μM as per our metabolomics data for L82V (1.38 fold of WT levels). With these values, $(V_{mut} - V_{WT})$ showed a significant correlation (Spearman $\rho = -0.54, p = 2.8e-2$) with observed change in lag times of mutants in the absence of any additional AMP in the medium (Fig 5C).

To gain further insight into the regimes of intracellular AMP concentrations in mutants and WT that may lead to significant correlation between flux and lag times, we modeled intracellular AMP concentration for WT in stationary phase to vary from 25 μM to 100 μM, which are ~10-35% of AMP during the exponential phase (280 μM). For each concentration of $[AMP]_{WT}$, we also assumed the ratio of $[AMP]_{mut}/[AMP]_{WT}$ to vary from 0.4 to 4.0. For each pair of $[AMP]_{WT}$ and $[AMP]_{mut}$, we calculated the Spearman correlation coefficient between $(V_{mut} - V_{WT})$ and experimentally observed change in lag times of mutants (at 0 external AMP) as in Fig 5C. Interestingly, we find that the correlation is significant only if mutants have higher AMP levels than WT during the stationary phase (Fig 5D).

Overall, these results show that even under physiological conditions substrate inhibition is essential for mutant Adk activity and it can cause loss of fitness for such mutants. This in turn can act as an evolutionary constraint that limits excessive protein stability for adk.

## Discussion

The physical or evolutionary reasons behind relatively low stability of modern-day mesophilic proteins have been at the center of a long-standing debate. Theoretical studies explain this based on the large supply of destabilizing mutations. A competing hypothesis suggests fitness penalty at high stability, however no experimental evidence exists till date. Here, we engineered stabilized mutants of an essential *E. coli* enzyme Adenylate Kinase and show that though such mutants have improved catalytic efficiency in terms of $k_{cat}/K_M$, they also exhibit strong substrate inhibition by AMP. AMP substrate inhibition is a well-known phenomenon for *E. coli* Adk (27, 28), here we uncover that this property of the enzyme is modulated by stability. Remarkably, destabilized mutants of Adk are also significantly less inhibited by AMP, to an extent that it is completely abolished for some mutants. We also show that the substrate inhibition constant $K_I$ shows a trade-off with enzyme efficiency $k_{cat}/K_M$. This observation implies that while improving stability that lead to more efficient Adks, the net velocity given by equation (1) will always be limited by substrate inhibition in the regime of high substrate concentrations.

We also show that substrate inhibition can result in pronounced fitness effects. In the presence of excess AMP, we show that the observed fitness effects can be accurately described using flux dynamics theory and a biophysical fitness function. More interestingly, variations in fitness effects were also observed under physiological conditions in the absence of any additional AMP. Using metabolomics data, we were able to explain this variation based on differential levels of AMP in the stationary phase for inhibition-prone mutants. This result has important implications in terms of evolution of protein stability. Due to absence of any evidence of fitness penalty at high stability, it was always believed that the fitness landscape is monotonic with respect to protein stability: reduced fitness at low stability due to low folded fraction, and essentially reaching a plateau once the fraction unfolded becomes negligible. Our results show that this landscape can be non-monotonic for some proteins, where high stability can impair activity through substrate inhibition. Our findings can be depicted in a schematic fitness landscape as shown in Fig 6, where a bell-shaped fitness landscape along stability axis arises due to substrate inhibition at high stability and it may indicate origins of moderate stabilities in Adk. Since substrate inhibition is a reality for ~20-30% of natural enzymes (38, 39), the observed phenomenon of increased inhibition upon stabilization might be applicable to these enzymes as well.

Of course, the fitness penalty at high stability and the relief of substrate inhibition upon destabilization does not suggest that large destabilization is beneficial for the enzyme. Destabilization concomitantly worsens $K_M$, and reduces intracellular abundance of the enzyme through greater contribution of degradation in turnover (32), as seen here and also in our previous study (23). Together this causes reduced flux through the enzyme and ultimately results in increase in lag times and lower fitness, even in the absence of AMP inhibition.

An interesting observation from our study is that $K_I$ of WT (~900 μM) is much higher than intracellular AMP concentration (280 μM), which implies that substrate inhibition is effectively not realized for WT under physiological conditions. On the contrary, for several stabilizing mutants the $K_I$ values are in the range of 200-300 μM (Table 1). Since intracellular metabolite concentrations are generally tightly regulated (40), a reasonable evolutionary strategy would be to evolve Adk stabilities in a range where inhibition effects are minimal.

We found that higher stability also leads to tighter binding to AMP $\left(K_M\right)$ and lower $k_{cat}$. Though the mechanism behind this is unclear, it is possible that stabilizing mutants stabilize the ligand bound closed-state of Adk more than the unbound open state (Fig S3), and hence increase the free energy of binding to the ligand (improved $K_D$, hence improved $K_M$). However, such effect might also decrease $k_{cat}$ by increasing the activation barrier between the ligand-bound closed state and the transition state of the phosphate transfer reaction. The stabilization may also result in lower interconversion rate between open and closed states in stable mutants. This is in agreement with recent single-molecule FRET study in Adk where mutations that reduced the rate of interconversion also reduced the $k_{cat}$ (41).

Of course, the big question remains unanswered: why does higher stability cause stronger inhibition? At the heart of this, lies the mechanism of AMP substrate inhibition of Adk, which has been an area of long-standing research. The general mechanism of substrate inhibition is assumed to be uncompetitive where AMP binds at an independent allosteric site (27). An alternative mechanism is that binding of AMP to its own site causes closure of the ATP binding pocket, leading to inhibition (28). A third mechanism is that inhibition is due to AMP binding competitively at the ATP-binding site (42). Though elucidation of the exact mechanism is beyond the scope of this work, this knowledge will be crucial to understand how stability modulates inhibition. We like to note in passing that in our attempt to get a mechanistic insight, we measured binding affinities of WT Adk, mutants M21A+L107I and L82V to the inhibitor Ap5A, which binds to both AMP and ATP binding site simultaneously (43). In accordance with the trends in $K_I$ values, $K_D$ for Ap5A were in the order of L82V< M21A+L107I <WT, implying that mutants that bind Ap5A strongly are also the ones that show strong AMP inhibition (Fig S4). This might hint at the third mechanism, in which stabilization of Adk somehow improves affinity of AMP at the ATP binding site, however further experiments are required to completely understand the mechanism.

# Methods

<u>Selection of mutations:</u> We attempted to design stable mutant of Adk with as few substitutions as possible. It is known that consensus mutations can increase the protein stability (44-48). Hence, we built a dataset of 895 adk protein sequences collated from ExPASy Enzymes (49) (as of Nov. 2012), clustered them at 99% sequence identity using CD-HIT (50), aligned and counted frequency of each amino acid and gaps at every position. Consensus amino acid at a position is the one with the highest frequency. In *E. coli* Adk, 56 positions were found to be out of consensus. Further pruning was done based on following criteria: a residue whose side chain is not involved in any hydrogen bonding and is at least 8Å away from bound inhibitor Ap5A based on pdb 1ake (26). Structure of *E. coli* Adk can be divided into three domains: LID (residues 118-160), NMP (residues 30-67), and Core (residues 1-29, 68-117, and 161-214). There are 28 residues which satisfy abovementioned criteria, of which 20 are in the Core domain, 5 in the LID, and 3 in the NMP. We randomly chose 6 positions from Core domain and constructed individual back-to-consensus mutations: M21A, M96L, L107I, V169E, L209I, and E210L. We define the active-site as the residues whose accessible surface area changes by more than 5 Å$^2$ in the presence of the inhibitor Ap5A. A similar criterion was used to define the residues contacting the active site. All the 6 selected positions were not only away from the inhibitor, but also not in direct contact with any active site residues.

Q16L and T179M were previously found to be stabilizing in Adk of *Bacillus subtilis* (24, 25). Based on that, we constructed Q16F, Q16Y, and T175M in *E. coli* Adk at structurally equivalent positions to Bacillus Adk. Such positions were determined by aligning structures of Adk from *E. coli* (pdb: 1ake) and *B. subtilis* (pdb: 1p3j (51)) using MUSTANG (52).

Additionally, we combined individually-stabilizing mutations to make 2- or 3-site mutants if their C$_\alpha$ atoms are at least 10 Å far apart from each other.

For destabilizing candidates, we chose several mutants from our previous study (23).

<u>Mutagenesis and protein purification:</u> We cloned *adk* gene in pET28a(+) vector between *Nde*I and *Xho*I restriction sites. The mutagenesis was carried out by amplifying the whole plasmid using inverse PCR protocol, KOD hot-start DNA polymerase, and a pair of partially complementary mutagenic primers (30-35 bp long). Such amplified plasmids were transformed in *E. coli* DH5α competent cells for faithful propagation and storage. For protein purification, pET28a(+) plasmids with WT and mutant *adk* were transformed in *E. coli* BL21(DE3), grown in 1 liter terrific broth and induced with 1 mM IPTG at $OD_{600}$ of 0.6. The proteins were purified using Ni-NTA affinity columns (Qiagen) and subsequently passed through a HiLoad Superdex 75 pg column (GE). The proteins eluted as a monomer. The corresponding fractions were pooled together, concentrated and dialysed against 10 mM potassium phosphate buffer (pH 7.2). The concentration of the proteins was measured by BCA assay (ThermoScientific) with BSA as standard.

Thermal denaturation: We used 20 µM of protein for assessing thermal stability of adk variants by differential scanning calorimetry (nanoDSC, TA instruments). The scans were carried out from 10 to 90 °C at a scan rate of 60 °C/hr. The thermodynamic parameters were derived by fitting the data to a two-state unfolding model using NanoAnalyze (TA instruments).

Enzyme activity: Adk catalyzes the following reaction: $ATP + AMP \xrightleftharpoons{ADK} 2\,ADP$. We measured the activity of Adk in the direction of ADP formation by a continuous assay. The reaction mixture contained a fixed concentration of ATP (1000 µM), varying concentration of AMP (from 0 to 500 or 1000 µM), 5 mM $MgCl_2$, 65 mM KCl, 350 µM phosphoenolpyruvate (PEP), and 300 µM of NADH. The mix was incubated at 25 °C for 5 minutes for equilibration. The reaction was initiated by addition of 5 nM Adk (final concentration) and 2 units of pyruvate/lactate dehydrogenase mix. The kinetic traces were recorded every 5 s for a total time of 2 minutes. The initial rates $(v_0)$ were estimated and plotted against AMP concentrations. As observed previously (27, 28), the kinetic data for varying AMP exhibited a signature of substrate inhibition and we modeled it assuming uncompetitive mode of inhibition as follows:

$$v_0 = \frac{k_{cat}[E_0][AMP]}{K_M + [AMP]\left(1 + \frac{[AMP]}{K_I}\right)} \qquad (4)$$

where $K_I$ is the AMP inhibition constant, $K_M$ is Michaelis constant for AMP, and $E_0$ is the concentration of Adk used in the assay. In few cases where inhibition was not apparent, the data exhibited hyperbolic pattern. Such traces did not fit well to equation (4) which was exemplified by large errors in $K_I$ compared to its mean fitted value. These cases were fitted by regular Michaelis-Menten equation:

$$v_0 = \frac{k_{cat}[E_0][AMP]}{K_M + [AMP]} \tag{5}$$

Generation of mutant strains: We generated the WT and mutant *adk* strains of *E. coli* MG1655 as described previously (23, 53). In short, *adk* variants were cloned in pKD13 having following cassette: htpG – REPt44 – Cam$^R$ – adk – REPt45 – Kan$^R$ – hemH. As indicated, chloramphenicol- and kanamycin-resistance genes are placed on either side of the *adk* gene, and long homology segments were added from upsteam and downstream genes. We amplified the whole cassette with ~800 bp of homology-segments, and electroporated in competent BW25113 cells in which λ-red system was already induced. The cells were recovered in 1 ml terrific broth for overnight at room temperature and eventually spread on LB-agar plates containing 34 µg/ml chloramphenicol and 50 µg/ml kanamycin. The colonies were sequenced for correct mutations. The mutant adk segments were subsequently transferred to *E. coli* MG1655 by P1 transduction, selected on LB-agar plates with both antibiotics as mentioned above, and the mutations were confirmed by sequencing.

Growth curve measurements and media conditions: The Adk strains were grown for 20 h at 30 °C from single colonies in M9 media (1× M9 salts, 1 mM MgSO$_4$, 0.2 % glucose). These primary cultures were normalized to a final OD$_{600}$ of 0.01 in fresh M9 media and the growth curves were monitored from three colonies in triplicates using Bioscreen C at 37 °C with data acquisition at every 15 min. For experiments with AMP, primary cultures were grown as mentioned above, and the secondary cultures were grown in M9 media with desired concentration of AMP from time 0.

We derived the growth parameters by fitting ln(OD) versus time plot (with $OD_{600} \geq 0.02$) with the following four-parameter Gompertz function as described previously (23):

$$\ln(OD) = \ln(OD_0) + \ln(K)\exp\left[-\exp\left(-\frac{t-\lambda}{b}\right)\right] \tag{6}$$

where $K$ is the fold-increase over initial population at saturation, the maximum growth rate is $\mu = \ln(K)/(b \cdot \exp(1))$, and the lag time $\lambda$ is the time taken to achieve the maximum growth rate. The error in parameters from replicates was found to be between 2-3% on an average, and it did not improve significantly upon increase in number of replicates.

Intracellular protein abundance: The WT and mutant strains in MG1655 were initially grown at 30 °C for 20 h in M9 medium. These primary cultures were normalized to $OD_{600}$ of 0.01 in fresh M9 media and grown for 5 hours at 37 °C. The cells were harvested and subsequently lysed with 1× BugBuster (Novagen) and 25 U/ml of Benzonase. The cell lysate was divided in two parts: one was used to estimate the total amount of proteins, and the other was for the specific fraction of Adk. The total amount of proteins in cell lysate was estimated by BCA assay (ThermoFisher). We used SDS-PAGE followed by western blot for estimating the intracellular abundance of Adk. The Adk bands were detected using anti-Adk polyclonal antibodies custom-raised in rabbit (Pacific Immunology). The band intensities on western blot were quantitated using ImageJ and were further normalized by the total protein abundance in that lysate (estimated as mentioned above). We used three colonies to estimate the intracellular abundance of Adk variants.

Intracellular metabolite levels: The primary cultures of Adk variants were grown in M9 medium as discussed above. The cultures were normalized to $OD_{600}$ of 0.01 in fresh M9 and grown at 37 °C for 5 h for exponential phase, and for 12 and 20 h for early and late stationary phase, respectively. In case of samples with AMP, the primary cultures were diluted in M9 with 400 μM of AMP and grown at 37 °C for ~8 h. The culture volumes corresponding to ~$3 \times 10^9$ cells (assuming $OD_{600}$ of 1 ≈ $8 \times 10^8$ cells/ml) of Adk strains were mixed with ~$5 \times 10^8$ cells of WT grown in M9 containing uniformly 13C-labeled glucose (Cambridge Isotope Laboratories, Inc). The labeled culture was used to correct for variability introduced at the sample processing stage. The mixed cells were harvested, and the dry pellet weight was recorded. Subsequently the metabolites were extracted in 80:20 methanol:water and detected by LC-MS as described previously (54).

The 13C-labeled metabolites were detected using 5 ppm accuracy window around their predicted monoisotopic m/z value and retention time. The retention time for the labeled metabolites was same as that for the unlabeled metabolites. For correction, we used approximately 16-18 labeled metabolites that were common in all the samples. The log of peak area for the labeled metabolites was linearly regressed against data from the first colony of WT (arbitrary choice of reference) and the regression parameters (slope and intercept) were used to correct the unlabeled peak areas.

## Acknowledgements

This work was supported by NIH R01 GM068670 to E.I.S.

**Figure Legends**

**Fig 1: Protein stability modulates substrate inhibition by AMP.** (A) Crystal structure of Adenylate Kinase from *E. coli* (PDB ID 4ake). The core domain is colored in green, while the LID and NMP domains are shown in gray. The $C_\alpha$ atoms of active-site residues are shown in pink, and the blue spheres represent the 7 positions which were mutated in this study. Q16 is within 4Å of active site and is shown in a red sphere. (B) Representative enzyme activity curves of WT, destabilized mutant V106W and a stabilized triple mutant M21A + V169E + L209I as a function of varying AMP concentrations. While WT shows moderate drop in velocity of the reaction at high AMP concentration, it becomes much more pronounced for the triple mutant, resulting in stronger $K_I$ (Table 1). Data for both WT and the triple mutant were fitted to equation (1) to derive activity parameters and are shown as solid lines. Destabilized mutant V106W does not show detectable inhibition in the range of AMP concentration studied, and hence was fitted with the conventional Michaelis-Menten equation (also see Methods). (C) Inhibition constant $(K_I)$ derived using equation (1) shows trade-off with stability. WT Adk is shown in green, while L82V is shown in light blue. Pearson correlation was calculated between $\Delta T_m$ and $\log(K_I)$. The correlation values with and without L82V are $r = -0.67$, $p = 1.1\mathrm{e}{-3}$, and $r = -0.80$, $p = 3.3\mathrm{e}{-5}$, respectively. Error bars in (B) and (C) are SEM of at least three repeats.

**Fig 2: Correlation of activity parameters with protein stability.** (A) $k_{cat}/K_M$ correlates positively with stability. WT is shown in green, and L82V in light blue. In pink are shown proteins that contain mutations at position Q16, which is very close to the active-site ($C_\alpha$ is < 4 Å of active site residues). Q16 mutants were not considered while calculating correlation coefficient. (B) $K_I$ of mutant proteins positively correlates with $K_M$ of AMP, implying that mutants that bind strongly to AMP also exhibit stronger AMP inhibition. Pearson correlation was calculated using log values of both $K_M$ and $K_I$. (C) $K_I$ negatively correlates with $k_{cat}/K_M$ of mutant proteins. This correlation is primarily driven by positive correlation between $K_M$ and $K_I$. Pearson correlations calculated using log values of $K_I$ are $r = -0.65$, $p = 2.8\mathrm{e}{-3}$, and $r = -0.55$, $p = 0.03$ without and with mutant L82V (shown in light blue).

**Fig 3: Fitness effects in the presence of excess AMP.** Growth rate (A) and lag times (B) of mutant Adk strains in the absence (0 μM) and presence of different concentrations of AMP (50-400 μM) in growth media. Except for L82F, most mutants with weaker inhibition than WT show minimal changes in growth rate and lag times upon exposure to AMP. On the other hand, mutants with stronger $K_I$ than WT generally exhibit considerable drop in fitness with increasing concentrations of AMP. The stability of the mutants is shown in (C). (D) Change in growth rate of Adk mutants in the presence of external AMP relative to zero AMP is plotted against change in flux $(V_{AMP} - V_0)$ in those conditions. The flux is calculated using equation (3). (E) Similar plot as in (D) for change in lag times. In both (C) and (D), the Spearman correlation coefficient is highly significant.

**Fig 4: Mutants with stronger inhibition accumulate all three adenylate species in the presence of excess AMP during exponential phase.** In the absence of AMP, intracellular levels of ATP, ADP, AMP are similar in WT and mutants V106W, L107I+V169E and L82V. However, in the presence of 400 μM external AMP, mutants L107I+V169E and L82V which show strong AMP inhibition, show considerable accumulation of all three adenylate species. In comparison, mutant V106W which shows no AMP inhibition, does not show measurable accumulation. WT behaves similar to V106W. In all cases however, cells convert AMP (a putative starvation signal) to IMP, which leads to increase in intracellular IMP levels.

**Fig 5: Fitness effects under physiological conditions.** (A) The metabolites were measured in the absence of any additional AMP at 5, 12, and 20 h of growth, timepoints that correspond to the exponential phase, and early- and late-stationary phase, respectively. Levels of adenylate metabolites drop drastically in stationary phase compared to that in the exponential phase. The overall trend of drop is similar in both WT and L82V. ATP for WT at 12 h was not detected faithfully. (B) Levels of all three adenylate metabolites is higher in L82V relative to WT during stationary phase. (C) Change in lag times of mutant Adk strains relative to WT at zero external AMP is plotted against change in flux $(V_{mut} - V_{WT})$ considering $[AMP]_{WT}$ as 70 μM (25% of that in exponential phase, 280 μM) and $[AMP]_{mut}$ as 96 μM (1.38-fold over WT levels). The Spearman correlation coefficient is significant $\rho = -0.54$, $p = 2.8e-2$. (D) *Modeling of AMP regimes for WT and mutants:* Intracellular concentration of AMP in WT at stationary phase was assumed to vary from 25 μM to 100 μM at an interval of 1 μM. At each concentration of WT, the ratio of $[AMP]_{mut}/[AMP]_{WT}$ was varied from 0.4 to 4.0 at an interval of 0.01, and change in flux $(V_{mut} - V_{WT})$ was calculated as in (C). We assumed all mutants contain same amount of AMP in the stationary phase. The plot shows a contour map of the spearman correlation coefficients between change in lag times of Adk variants at zero external AMP and $(V_{mut} - V_{WT})$, calculated for different pairs of $[AMP]_{WT}$ and $[AMP]_{mut}$. Red region corresponds to the highest values of correlation while blue represents lowest. The dashed yellow line represents the contour line of $p = 0.05$, above which the correlation is significant, i.e., $p < 0.05$. The blue dashed lines represent AMP concentrations used in panel (C). This plot shows that flux through mutant Adk can describe the variation in lag times significantly only when mutants have higher concentration of AMP in stationary phase relative to WT. Our experimental data with mutant L82V is in agreement with this finding.

**Fig 6: Schematic fitness landscape depicting purifying selection at high stability.** A schematic landscape depicting a fitness cost at high protein stability primarily arising through substrate inhibition. In Adk, increased stability of proteins results in increased substrate inhibition, which in turn reduces flux through the enzyme, and eventually causes a drop in fitness. Hence, substrate inhibition in Adk results in a bell-shaped fitness landscape along the stability axis signifying the purifying selection at high stability. The arrows on all axes are pointed to the increasing direction.

**Table 1: Biophysical properties of Adk mutants**

Fig 1

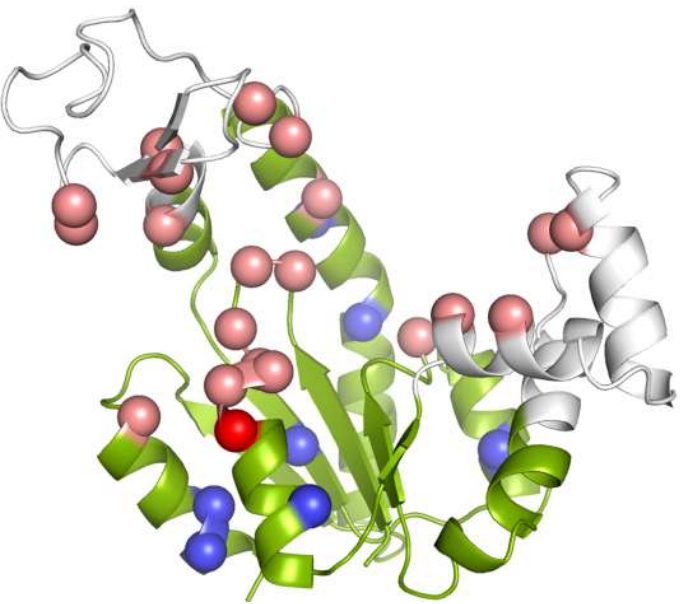
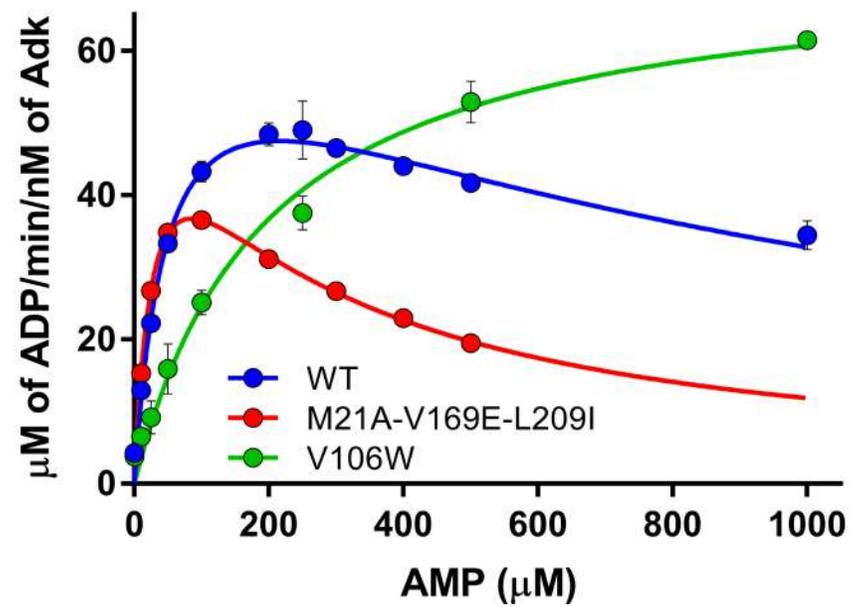
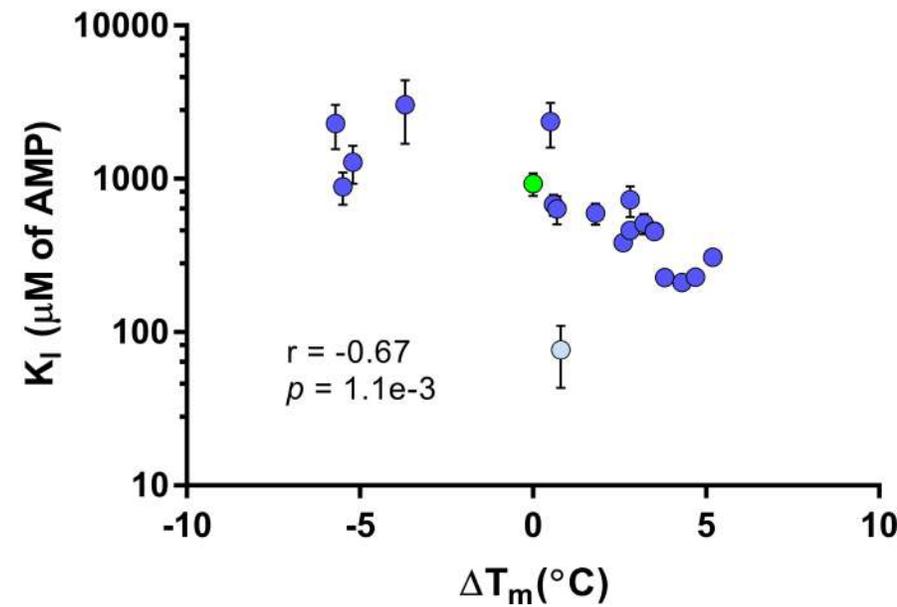

Fig 2

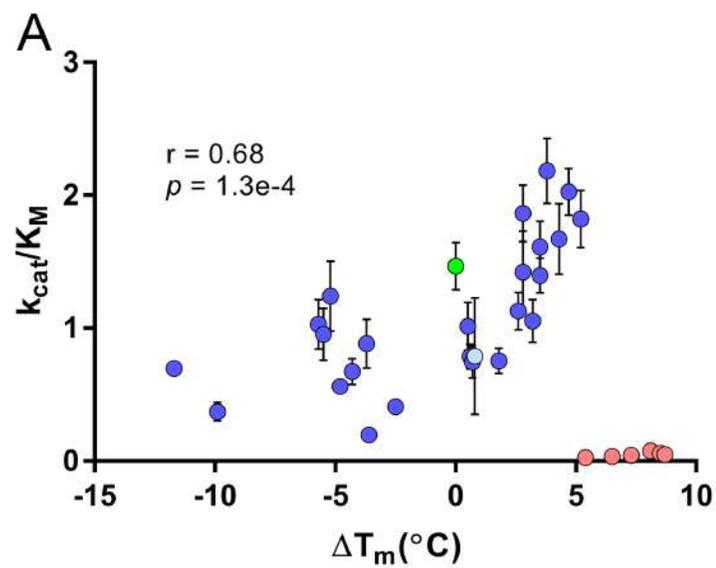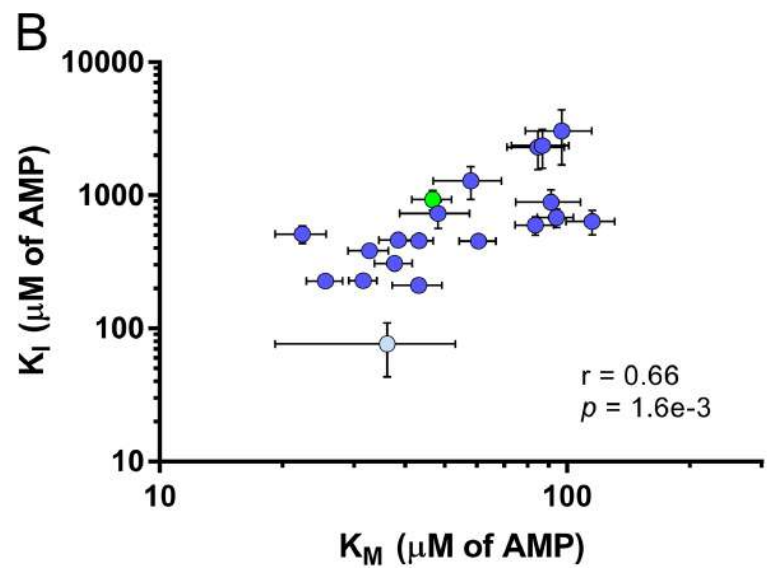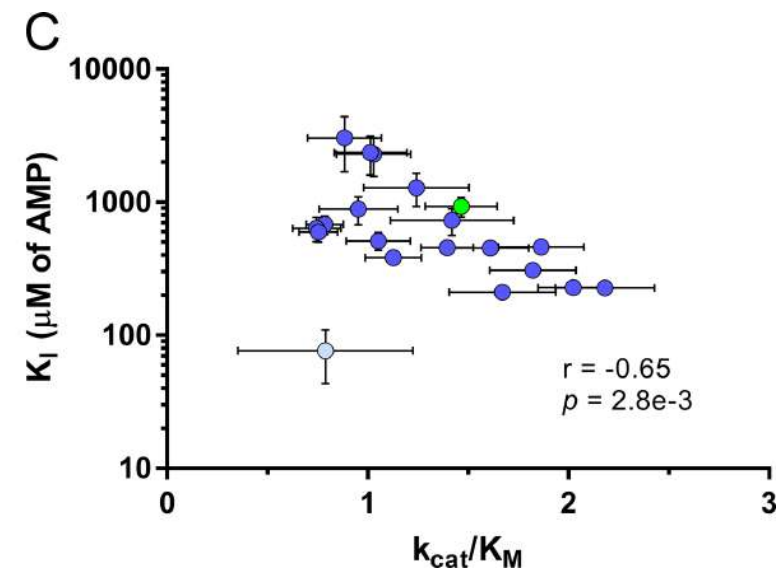

Fig 3

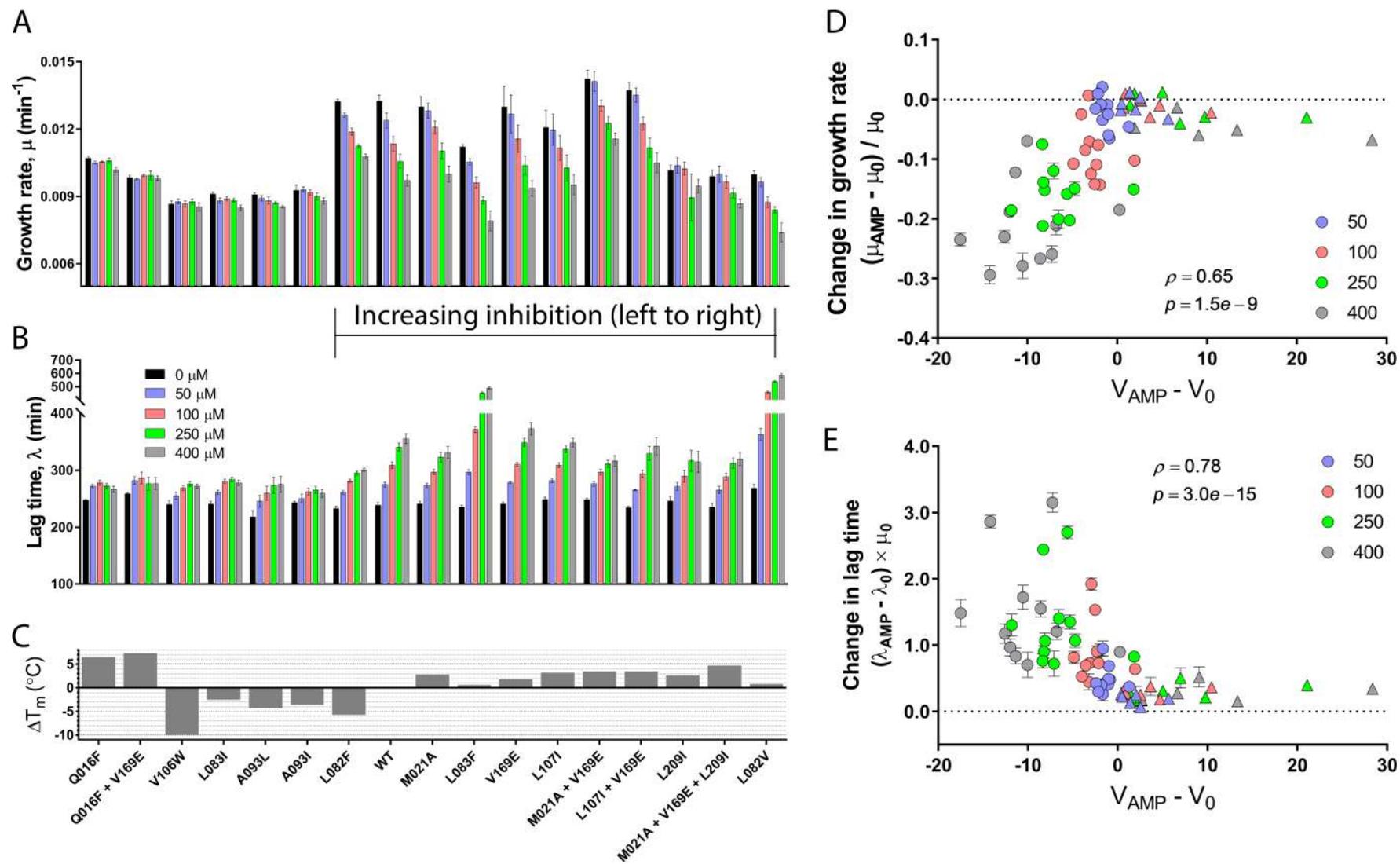

Fig 4

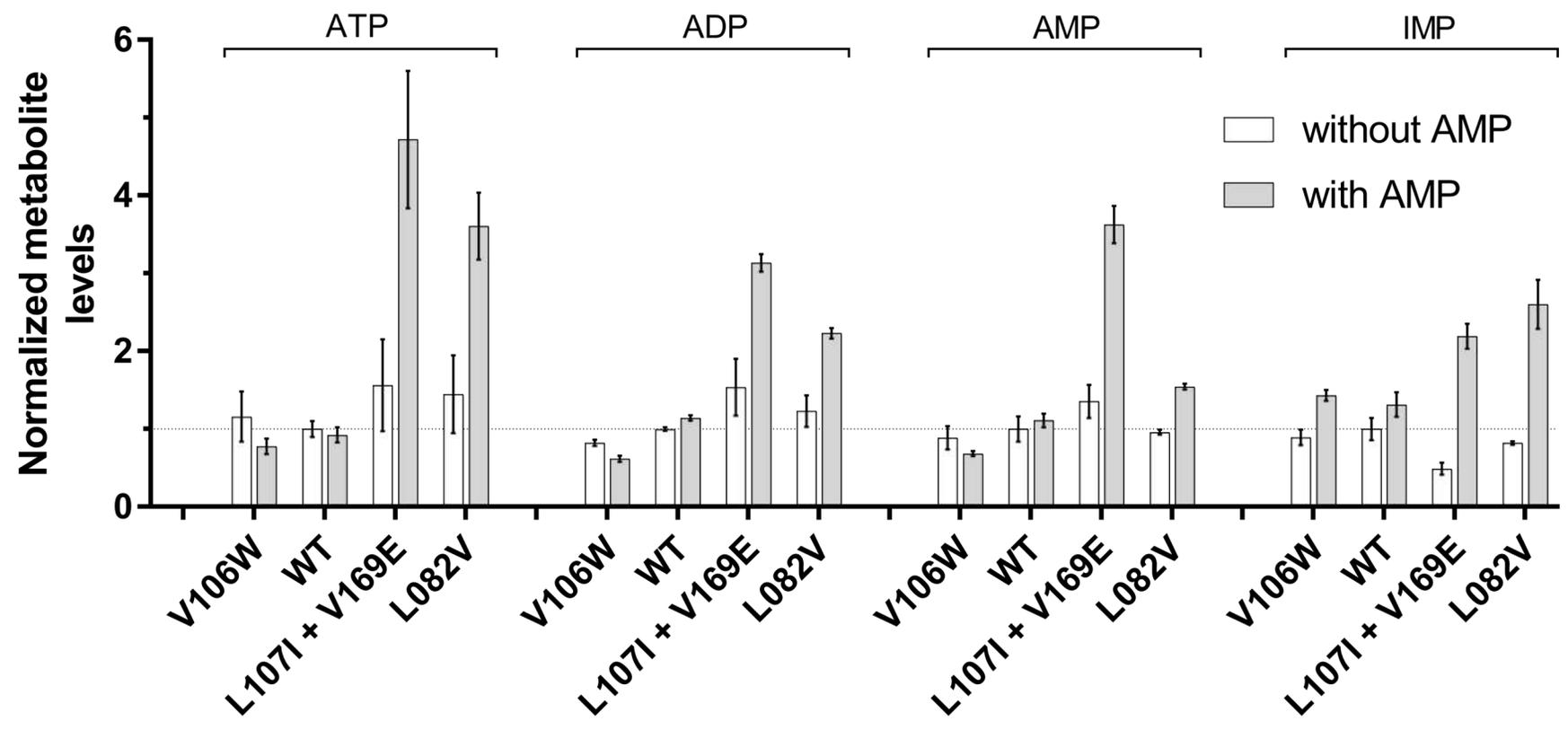

Fig 5

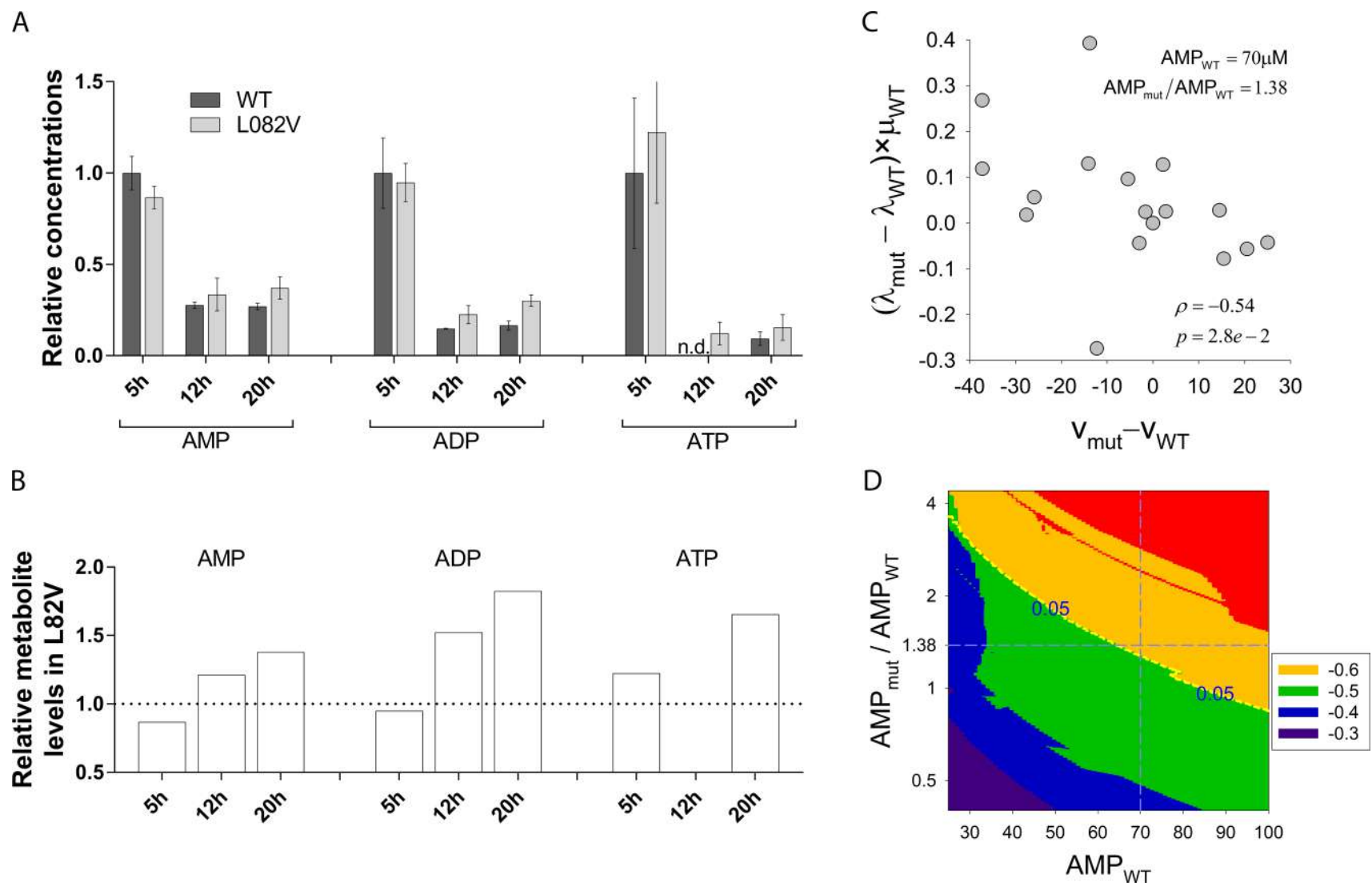

Fig 6

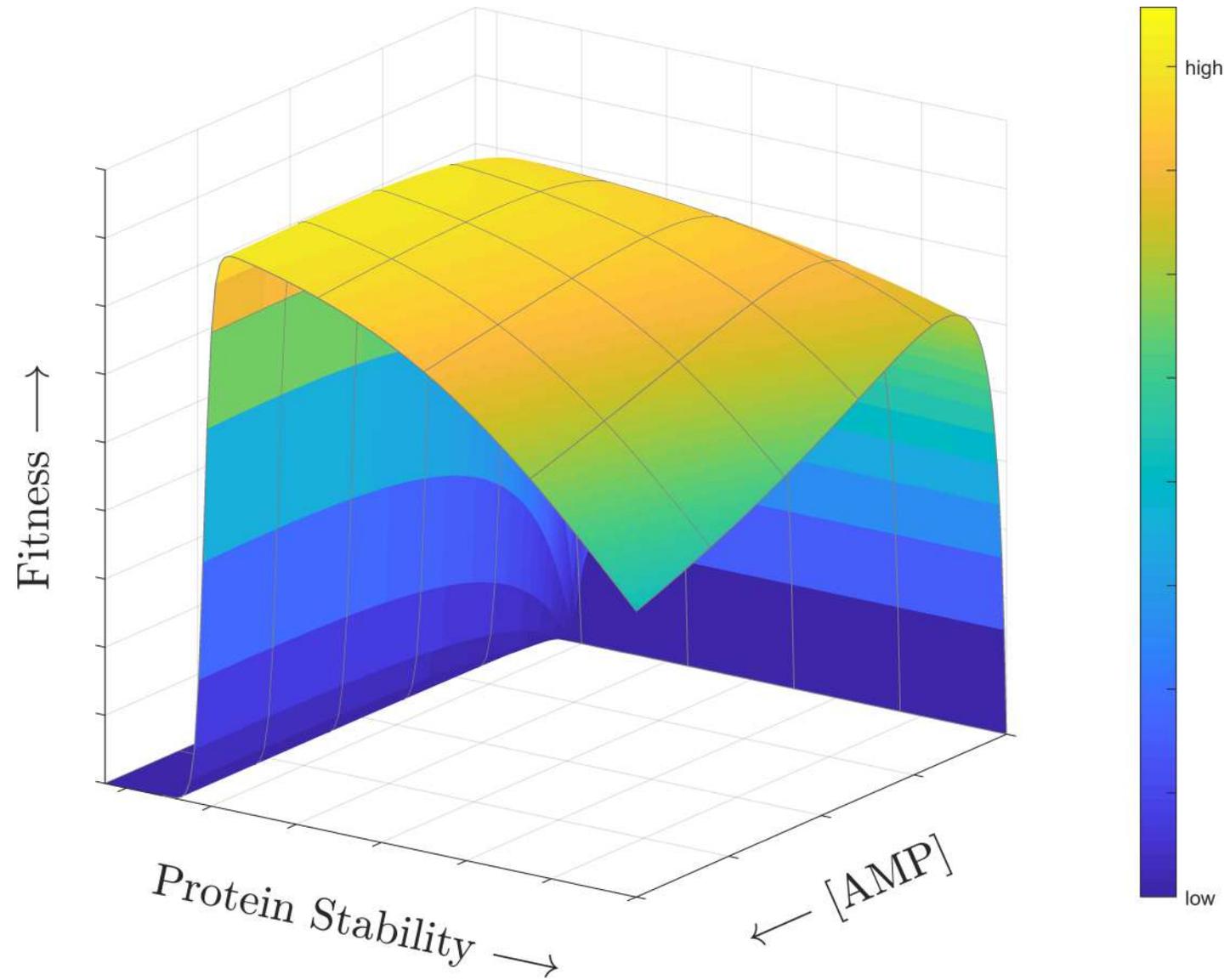

**Table 1: Biophysical propeties of Adk mutants**

| Mutant | $\Delta T_m$ (°C)[a] | $k_{cat}$ (μM of ADP/min/nM of Adk)[b] | $K_M$ (μM of AMP)[b] | $K_I$ (μM of AMP)[b,c] | distance from AP5A (Å) |
|---|---|---|---|---|---|
| V106H | -11.7 | 93.2 (2.1) | 133.6 (9.5) | NA[d] | 13.4 |
| V106W | -9.9 | 72.6 (4.6) | 195.7 (35.3) | NA[d] | 13.4 |
| L082F | -5.7 | 87.3 (6.9) | 84.8 (13.7) | 2302.0 (738.2) | 10.5 |
| A093F | -5.5 | 87.0 (8.6) | 91.3 (16.5) | 889.1 (209.7) | 8.7 |
| V106A | -5.2 | 72.0 (6.5) | 58.0 (11.1) | 1287.0 (355.5) | 13.4 |
| V106L | -4.8 | 106.6 (2.4) | 190.5 (12.2) | NA[d] | 13.4 |
| A093L | -4.3 | 80.8 (3.4) | 119.8 (16.5) | NA[d] | 8.7 |
| A093Y | -3.7 | 85.7 (7.9) | 97.0 (18.0) | 3046.0 (1349.0) | 8.7 |
| A093I | -3.6 | 69.0 (5.6) | 349.8 (67.8) | NA[d] | 8.7 |
| L083I | -2.5 | 105.9 (5.4) | 259.1 (34.7) | NA[d] | 8.8 |
| WT | 0.0 | 68.6 (3.6) | 46.8 (5.2) | 930.5 (154.7) | |
| Y182F | 0.5 | 88.1 (7.0) | 87.0 (14.0) | 2365.0 (766.8) | 10.7 |
| L083F | 0.6 | 73.9 (4.5) | 94.1 (9.6) | 680.2 (106.1) | 8.8 |
| E210L | 0.7 | 85.8 (7.3) | 115.1 (15.7) | 638.1 (132.1) | 13.1 |
| L082V | 0.8 | 28.5 (8.3) | 36.2 (17.0) | 76.6 (33.4) | 10.5 |
| V169E | 1.8 | 63.1 (4.0) | 83.6 (9.2) | 595.9 (92.9) | 10.0 |
| L209I | 2.6 | 36.9 (2.0) | 32.7 (3.7) | 382.9 (45.0) | 9.4 |
| M021A | 2.8 | 68.5 (6.4) | 48.2 (9.4) | 729.5 (166.3) | 12.5 |
| V169E + L209I | 2.8 | 71.8 (3.6) | 38.5 (4.0) | 460.5 (54.2) | |
| L107I | 3.2 | 23.6 (1.4) | 22.4 (3.2) | 511.7 (77.4) | 9.5 |
| M021A + V169E | 3.5 | 60.4 (2.5) | 43.3 (3.6) | 456.4 (44.3) | |
| L107I + V169E | 3.5 | 97.6 (5.6) | 60.6 (6.3) | 453.0 (57.6) | |
| M021A + L209I | 3.8 | 55.7 (2.7) | 25.5 (2.6) | 227.4 (20.7) | |
| M021A + L107I | 4.3 | 72.2 (5.6) | 43.2 (6.0) | 211.3 (28.1) | |
| M021A + V169E + L209I | 4.7 | 64.0 (2.5) | 31.6 (2.5) | 228.9 (16.4) | |
| M021A + L107I + V169E | 5.2 | 68.8 (3.7) | 37.7 (4.0) | 308.6 (32.9) | |
| Q016Y | 5.4 | 4.9 (1.1) | 186.3 (118.3) | NA[d] | 5.1 |
| Q016F | 6.5 | 10.6 (1.4) | 324.2 (101.0) | NA[d] | 5.1 |
| Q016F + V169E | 7.3 | 20.6 (3.9) | 479.3 (156.5) | NA[d] | |
| Q016F + L107I | 8.1 | 21.9 (2.7) | 286.8 (74.2) | NA[d] | |
| Q016F + L107I + V169E | 8.5 | 14.0 (2.8) | 241.0 (107.0) | NA[d] | |
| Q016F + M021A | 8.7 | 24.0 (3.4) | 491.7 (118.4) | NA[d] | |

a Tm of WT Adk is 54.9 °C
b numbers in the parenthesis are s.e.m. of minimum 3 repeats
c The kinetic data was fit to uncompetitive model of inhibition (see Methods Eq. 4)
d The kinetic data was fit to standard Michael Menten equation (see Methods Eq. 5)

# Substrate inhibition imposes fitness penalty at high protein stability

Bharat V. Adkar[a], Sanchari Bhattacharyya[a], Amy I. Gilson[a], Wenli Zhang[a,b], Eugene I. Shakhnovich[a,1]

[a]Department of Chemistry and Chemical Biology, Harvard University, 12 Oxford St., Cambridge, MA 02138, USA

[b]State Key Laboratory of Food Science and Technology, Jiangnan University, Wuxi 214122, China

[1]Correspondence should be addressed to E.I.S. (shakhnovich@chemistry.harvard.edu)

Supplementary Figures and Table:

4 Figures and 1 Table

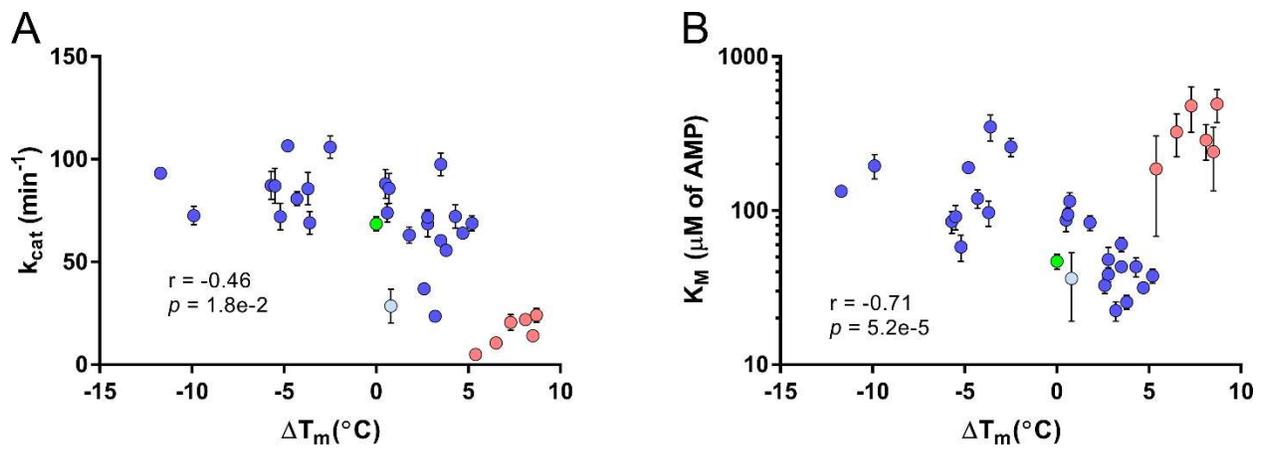

**Fig S1: Activity parameters vs stability.** Correlation between $k_{cat}$ (A) or $K_M$ (B) and protein stability is shown. Log values of $K_M$ were used for correlation calculations. The WT is shown in green, L82V in light blue, whereas all mutants involving mutation at Q16 position are shown in pink circles. The error bars are s.e.m. of three measurements. In both panels, Pearson correlations are calculated without considering Q16 mutants (red points).

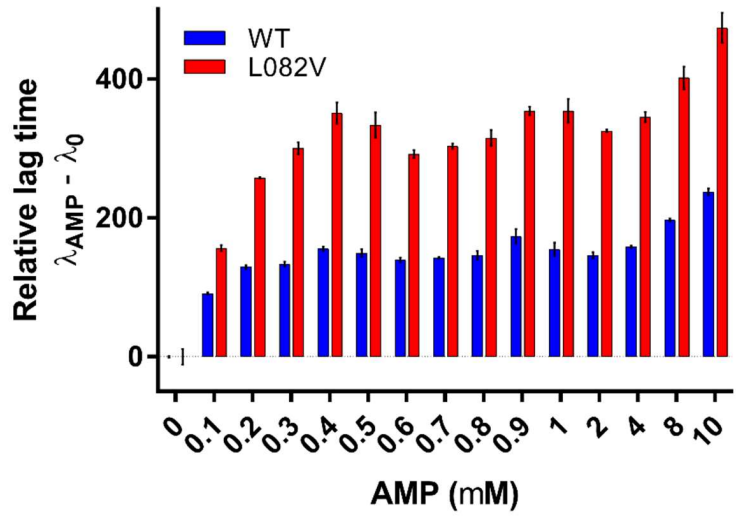

**Fig S2: Pilot experiment to estimate the dynamic range of AMP effect on fitness.** A pilot experiment with only WT and L82V, the most inhibited mutant in this study, were grown in M9 media containing various amounts of AMP. Relative lag times increase substantially up to 400 μM of AMP, following which the changes are smaller. The error bars are s.e.m. from three colonies.

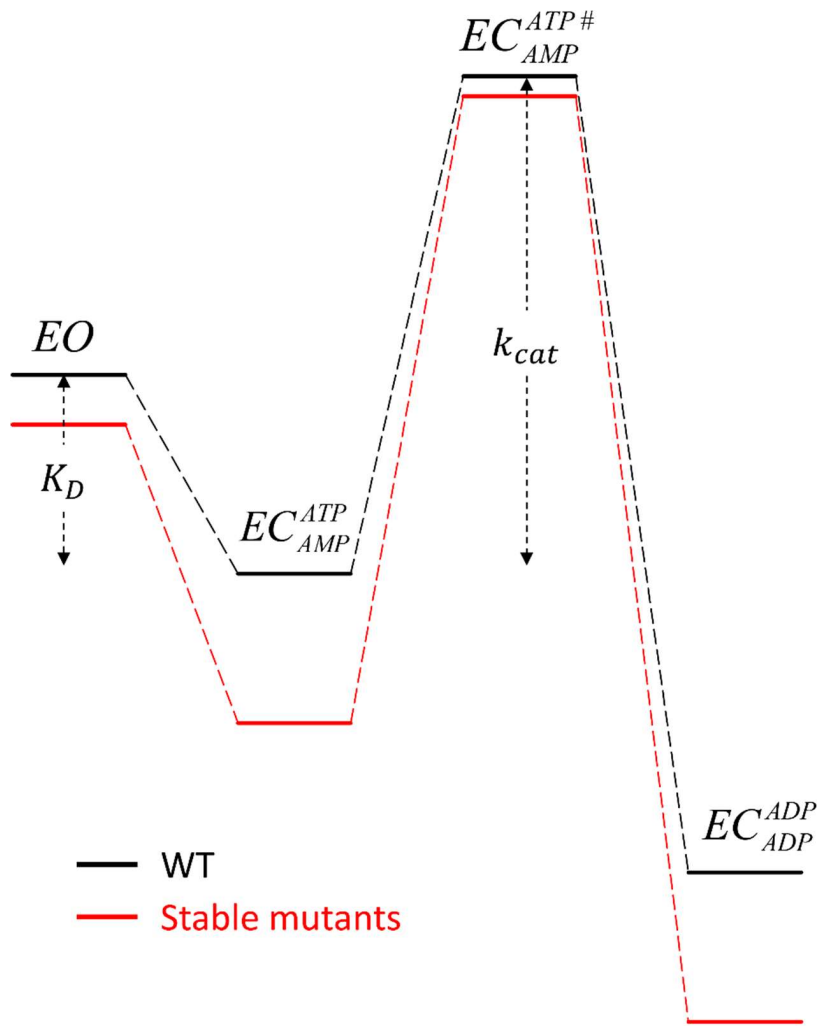

**Fig S3: Schematic of possible effects of stabilization on enzyme mechanics.** The schema depicts an energy landscape during catalysis. WT scheme is shown in black lines, whereas that of a stabilizing version is in red. EO and EC are the 'open' and 'closed' states of the enzyme. The more stable enzyme may preferentially stabilize the ligand-bound closed state more than the unbound open state, which will be reflected in stronger $K_D$ and hence stronger $K_M$. Such preference may also result in a high activation barrier $\left( EC_{AMP}^{ATP} \to EC_{AMP}^{ATP\,\#} \right)$, which results in reduced $k_{cat}$.

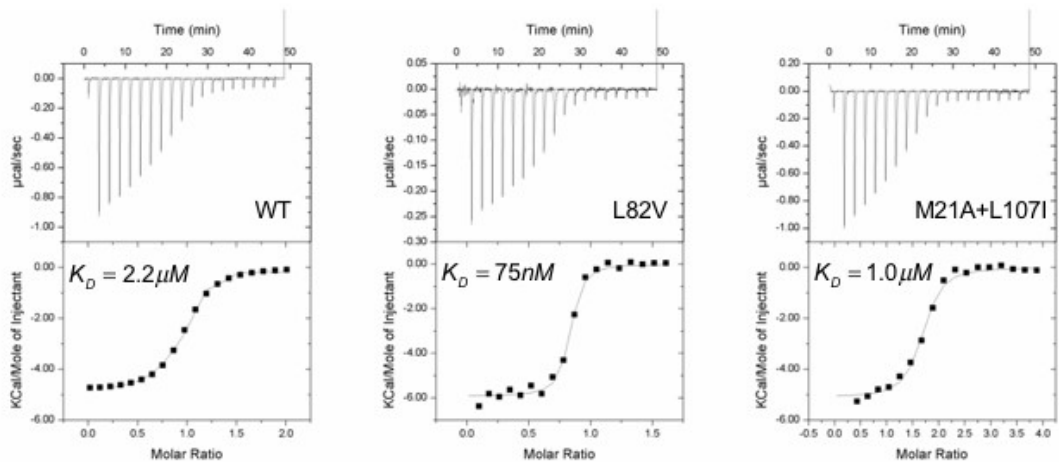

**Fig S4: Ap5A binding by ITC.** Binding of a bidentate inhibitor Ap5A was measured by ITC to WT, L82V, and M21A+L107I. The concentration of protein used in the cell was 96.7 µM of WT, and 24.1 µM each of the two mutants. 1 mM of the inhibitor was used in syringe for WT and M21A+L107I, whereas 200 µM was used for L82V titrations. The binding measurement was done at 25 °C.

**Table S1: Growth parameters of Adk strains**

| Mutant | Growth rate (min$^{-1}$) | | Lag time (min) | |
|---|---|---|---|---|
| | mean | SEM | mean | SEM |
| V106W | 0.0087 | 0.0001 | 240.4 | 6.1 |
| Q016F | 0.0107 | 0.0001 | 248.0 | 1.0 |
| Q016F + V169E | 0.0099 | 0.0001 | 259.3 | 1.3 |
| V106H | 0.0126 | 0.0001 | 189.7 | 2.4 |
| L083I | 0.0091 | 0.0001 | 240.9 | 4.2 |
| A093L | 0.0091 | 0.0001 | 218.4 | 10.2 |
| A093I | 0.0093 | 0.0002 | 243.3 | 2.4 |
| L082F | 0.0132 | 0.0001 | 233.2 | 3.5 |
| WT | 0.0133 | 0.0003 | 239.1 | 4.1 |
| M021A | 0.0130 | 0.0004 | 241.2 | 4.3 |
| L083F | 0.0112 | 0.0001 | 235.9 | 2.8 |
| V169E | 0.0130 | 0.0009 | 241.0 | 3.5 |
| L107I | 0.0121 | 0.0007 | 248.9 | 4.1 |
| M021A + V169E | 0.0143 | 0.0004 | 248.7 | 2.2 |
| L107I + V169E | 0.0137 | 0.0003 | 234.8 | 1.9 |
| L209I | 0.0102 | 0.0002 | 246.3 | 7.3 |
| M021A + V169E + L209I | 0.0099 | 0.0003 | 235.8 | 6.5 |
| L082V | 0.0100 | 0.0001 | 268.7 | 6.0 |

All experiments were done in triplicates with at least three colonies (biological repeats)